\documentclass[prl,twocolumn,showpacs,preprintnumbers,amsmath,amssymb,superscriptaddress]{revtex4}
\usepackage[english]{babel}
\usepackage{graphicx}
\usepackage{dcolumn}
\usepackage{epsfig}
\usepackage{setspace} 
\usepackage{bm}

\newcommand{\MET}{\mbox{$\protect \raisebox{.3ex}{$\not$}\!E_T$}}
\newcommand{\rmt}{\rm\textstyle}
\font\eightit=cmti8

\begin{document}
\pagenumbering{arabic}

\preprint{APS/123-TOP}

\title{Measurement of the $W$ Boson Polarization in Top Decay \\ at CDF at $\sqrt{s}$ = 1.8 TeV}

\normalsize    

\author{
D.~Acosta,$^{14}$ T.~Affolder,$^{7}$ M.G.~Albrow,$^{13}$ D.~Ambrose,$^{36}$   
D.~Amidei,$^{27}$ K.~Anikeev,$^{26}$ J.~Antos,$^{1}$ 
G.~Apollinari,$^{13}$ T.~Arisawa,$^{50}$ A.~Artikov,$^{11}$
% ...
}


\affiliation{
$^{1}$  {\eightit Institute of Physics, Academia Sinica, Taipei, Taiwan 11529, Republic of China} \\
% ...
$^{7}$  {\eightit University of California at Santa Barbara, Santa Barbara, California 93106} \\ 
% ...
$^{11}$  {\eightit Joint Institute for Nuclear Research, RU-141980 Dubna, Russia} \\
% ...
$^{13}$ {\eightit Fermi National Accelerator Laboratory, Batavia, Illinois
60510} \\
$^{14}$ {\eightit University of Florida, Gainesville, Florida  32611} \\
% ...
$^{26}$ {\eightit Massachusetts Institute of Technology, Cambridge,
Massachusetts  02139} \\   
$^{27}$ {\eightit University of Michigan, Ann Arbor, Michigan 48109} \\
% ...
$^{36}$ {\eightit University of Pennsylvania, Philadelphia, 
        Pennsylvania 19104} \\   
% ...
$^{50}$ {\eightit Waseda University, Tokyo 169, Japan} \\
% ...
}

\date{\today}

\begin{abstract}
The polarization of the $W$ boson in $t \to W b$ decay is unambiguously
predicted by the Standard Model of electroweak interactions and is a powerful
test of our understanding of the $tbW$ vertex.  We measure this polarization
from the invariant mass of the $b$ quark from $t \to W b$ and the lepton from
$W \to l \nu$ whose momenta measure the $W$ decay angle and direction of
motion, respectively.  In this paper we present a measurement of the decay
rate ($f_{V+A}$) of the $W$ produced from the decay of the top quark in the
hypothesis of V$+$A structure of the $tWb$ vertex.  We find no evidence for
the non-standard V$+$A vertex and set a limit on $f_{V+A}$ $<$ 0.80 at 95\%
confidence level. By combining this result with a complementary observable in
the same data, we assign a limit on $f_{V+A}$ $<$ 0.61 at 95\% CL.  This
corresponds to a constraint on the right-handed helicity component of the W
polarization of $f_{+} < 0.18$ at 95\% CL.  This limit is the first
significant direct constraint on $f_{V+A}$ in top decay.
\end{abstract}

\pacs{14.65.Ha, 12.15.Ji, 12.60.Cn, 13.88.+e \hspace{2cm} FERMILAB-PUB-04-353} 

\maketitle


%
The large value of the top quark mass has led to speculation that the top
quark could play a role in the mechanism of the electroweak symmetry breaking
\cite{hill}.  If so, the electroweak interactions of the top quark could be
modified \cite{Peccei:kr}.  Such a modification could alter the V$-$A
structure of the $tbW$ interaction which in turn would lead to an altered $W$
polarization in top decay \cite{Kane:1991bg,Jezabek:1994zv,Nelson:1997xd}.
Possible scenarios that would introduce a V$+$A contribution to the $tbW$
vertex include $SU(2)_L \times SU(2)_R$ extensions of the standard model
\cite{review}.  One such model invokes new mirror particles to assist a
top-condensate in breaking electroweak symmetry \cite{triant}. The theory of
``beautiful mirror'' fermions predicts a fourth generation up-type quark with
right-handed weak interactions which could contaminate the top sample or
induce a right-handed top electroweak interaction by mixing with the top
quark
\cite{Choudhury:2001hs}.

Indirect limits of right-handed $t \to b W$
currents have been placed using the process $b \to s \gamma$,
which proceeds via an electroweak radiative penguin process \cite{cleo}.
These limits are stringent, but scenarios can be envisaged where other
contributions to $b \to s \gamma$ might invalidate these bounds.
The goal of this study is a direct measurement of the $tbW$ vertex
from the electroweak decay of top.

The spin-one $W$ has three possible helicities; for the $W^+$ we label
these as $-1$ (left-handed), 0 (longitudinal), and $+1$ (right-handed),
with the opposite convention for the $W^-$.  Because $M_t>M_W$, a
large fraction of the $W$ bosons produced in top decay will be
longitudinally polarized \cite{Kane:1991bg}.  The fraction is given by
\begin{equation}
F_0 = \frac{M_t^2/M_W^2}{(M_t^2/M_W^2+2)}.
\end{equation}
For the current values of $M_t= 174.3 \pm 5.1$~GeV and $M_W = 80.425 \pm
0.038$~GeV \cite{pdg_masses}, this corresponds to $F_{0}=0.70 \pm
0.01$. 
If there were a
non-standard model V$+$A contribution to the top decay vertex, such
contribution would not
decrease the branching ratio to longitudinal $W$ bosons but would
instead decrease the branching ratio to left-handed $W$ bosons, replacing
some of this rate with an enhanced right-handed component.  

Leptons from the decay of longitudinally polarized $W$ bosons have a
symmetric angular distribution of the form $1-(\cos\psi_{\ell}^{\star})^2$,
where $\psi_{\ell}^{\star}$ is defined as the angle in the $W$ rest frame
between the lepton and the boost vector ($\vec{\beta}$) from the top
rest frame to the $W$ rest frame.  Maximal parity violation in the
V$-$A electroweak theory predicts that the non-longitudinal $W$
helicity is purely left-handed in the limit of massless final
state fermions.  This creates an asymmetric angular distribution of
the form $(1-\cos\psi_{\ell}^{\star})^2$ \cite{Kane:1991bg}.  Due to angular
momentum conservation, even though the massive top quark may be left- or
right-handed, positively polarized $W^{+}$ bosons are not
possible since a massless $b$ quark must be left-handed.  A small right-handed
component (0.04 \%) of the form $(1+\cos\psi_{\ell}^{\star})^2$ results
when the mass of the $b$ quark is considered.  

\begin{figure}
\resizebox{0.9\columnwidth}{!}{\includegraphics{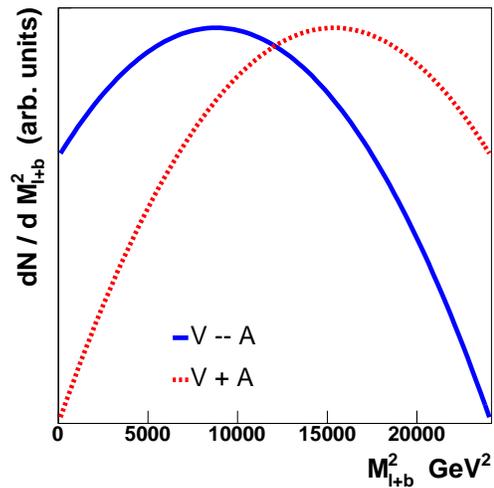}}
\caption{The theoretical distributions of $M_{\ell b}^2$ for purely V$-$A and
V$+$A hypotheses, using the correct lepton-$b$ pairing.  The $M_{\ell b}^2$ can
be used to discriminate between the two hypotheses as it peaks at higher
values for V$+$A. This ideal case does not include detector and trigger
effects or the intrinsic lepton-$b$ mass resolution.}
\label{fig:vma_vpa}
\end{figure}

This analysis exploits the relationship between the angle
$\psi_{\ell}^{\star}$ and the invariant mass of the $\ell b$ pair, produced in
the top decay chain $t \to W b$, $W \to \ell \nu$ to determine the polarization
of the $W$ boson.  
The angle $\psi_{\ell}^\star$ can be
related to the $\ell b$ invariant mass by
\begin{equation}
M_{\ell b}^2 = \frac{1}{2} (M_{t}^2 -  M_{W}^2)(1 + \cos\psi_\ell^\star).
\label{eq:m2cos}
\end{equation}

In the V$-$A theory, the lepton and $b$ jet in the $W$ rest
frame tend to move in the same direction, but in a V$+$A decay, the lepton
and $b$ jet typically move in opposite directions.  Therefore, $M_{\ell
b}^2$ would be larger on average from a V$+$A contribution as shown in
Fig.~\ref{fig:vma_vpa}.  This difference can be used to determine $f_{V+A}$,
the fraction of $t$ quarks which decay with a V$+$A interaction.  

If the interaction has both V$-$A and V$+$A contributions, the
total angular distribution will be approximately described by summing
over weighted linear combinations of the above angular distributions.
The summing of rates correctly describes the angular distribution from
longitudinal and either a pure V$+$A or V$-$A distribution; however, if
there is a combination of V$-$A and V$+$A interactions, they may 
interfere with some relative phase.
The present analysis neglects this interference, which would have the largest
impact for $f_{V+A} = 0.5$. These interference effects are only of
order $1/\gamma_b$, the boost of the $b$ quark in the top rest frame, and
therefore are estimated to affect the angular distributions \cite{Tim} at no
more than the 10\% level.  The associated uncertainty is therefore not
significant compared to expected statistical and systematic uncertainties.

Experimentally, $M_{\ell b}^2$ is a reliable observable in $t\bar{t}$ decay
at a hadron collider because no information about the top or $W$ rest frames
is required, and therefore the unknown boost of the $t\bar{t}$ system along
the beam direction does not disrupt the measurement.  This technique also
avoids the need to rely on the missing transverse energy ($\MET$) due to the
neutrino. The $\MET$ is poorly measured compared to other kinematic
quantities in the event and is ambiguous in events with two final state
neutrinos, e.g., both $W^+$ and $W^-$ from the $t\bar{t}$ decay leptonically.


The present study uses data from $p\bar{p}$ collisions at $\sqrt{s} = 1.8$
TeV collected by the Collider Detector at Fermilab (CDF)\cite{CDF} during the
period 1992-1995 (Run I). The integrated luminosity of the data sample is 109
$\pm$ 7 pb$^{-1}$.  Events were selected \cite{toplj,topdi} and assigned to
three different $t\bar{t}$ subsamples chosen for their low background and
high efficiency for $b$ jet identification.  Each sample is classified by the
number of leptons and identified $b$ jets in the final state.

The ``dilepton'' sample is dominated by $t\bar{t}$ in which both $W$ bosons
decay to an electron or muon and neutrinos.  Events are selected by
requiring $\MET >$ 25 GeV, one muon and one electron of opposite charge with
$P_T > 20$ GeV in the central pseudo-rapidity region ($|\eta| < 1.0$) \cite{eta}, and two jets with
$E_T > 10$ GeV and $|\eta| < 2.0$.  This is a subsample of the
dilepton events used in other analyses~\cite{topdi}, considering only $e +
\mu + jets$ events in order to remove the dominant background, 
which is Drell-Yan production of $ee$ or $\mu\mu$.
The significant remaining backgrounds are decays to electron and muon of $Z
\to \tau \tau$, $WW$ in association with extra jets, and $W$ production
associated with three or more jets, where one jet is misidentified as an
electron or a muon.  No attempt is made to identify $b$ jets explicitly.
However, initial and final state gluon radiation can result in extra jets, so
the $b$ jets are assumed to be the two highest $E_T$ jets, which is correct
in $\sim$80 \% of dilepton events.  There are four $M_{\ell b}$ combinations in
each dilepton event.

The other two samples used in the analysis require only one $W$ to decay into
an electron or muon and a neutrino and the other $W$ to decay hadronically
(``lepton+jets'').  These events are selected by requiring one electron or
muon with $P_T >$ 20 GeV, in the central region as above.  At least four jets
are required, three of which must have $E_T >$ 15 GeV, $|\eta| < 2.0$, and
the fourth must have $E_T >$ 8 GeV and $|\eta| < 2.4$. The background for
these events consists predominantly of direct production of a $W$ plus extra
jets and its behavior is modeled with the VECBOS generator \cite{VECBOS}.  To
reduce the background, at least one jet must be identified as a $b$ candidate
($b$-tagged) with a topological algorithm requiring tracks in the jet
reconstructed with the silicon vertex (SVX) detector to form a secondary
vertex \cite{toplj,oldtopresults}.  This requirement is 48 \%
efficient for tagging at least one $b$ jet in a $t\bar{t}$ event
\cite{cdf_mass_prd}.  Without any $b$-tag, the expected signal to background
ratio ($S/B$) of the sample is 0.4, whereas requiring one $b$-tag improves
$S/B$ to 5.3.  The $b$-tag also selects the jet to be paired with the lepton
to form $M_{\ell b}$.  Events with a single $b$-tagged jet comprise the
``single-tagged'' sample, and have one measured $M_{\ell b}$ which is correct
half the time. Events with both $b$ quarks tagged make up the
``double-tagged'' sample, have a $S/B$ of 24, and provide two $M_{\ell b}$
pairings, at least one of which combines the wrong $b$ with the $\ell$.

A total of 7 events were found in the dilepton $e \mu$ sample
with an expected background of 0.76 $\pm$ 0.21 events. In the
single-tagged sample 15 events were found with a background 2.0 $\pm$
0.7, and in the double-tagged sample there were 5 events with a 0.2
$\pm$ 0.2 background.  Note that since right-handed
leptons have higher $P_T$, an increase in events passing the lepton
$P_T$ trigger requirement could also indicate a V$+$A theory.  However,
any potential observed rate increase would be deemed to be {\em a posteriori}
knowledge from the point of view of this analysis, and therefore only the
shape of the $M^2_{\ell b}$ distributions is considered.


%

The $M_{\ell b}^2$ distributions of the data are fit to a linear combination of
three predicted $M_{\ell b}^2$ distributions: $t\bar{t}$ production with a V$-$A
interaction, $t\bar{t}$ production with a V$+$A interaction, and background.  The fit maximizes
a binned likelihood as a function of $f_{V+A}$.  
Likelihood scans are performed both inside and outside
the physical region of $[0,1]$ in $f_{V+A}$, and the level of
backgrounds in each fit is allowed to vary within the estimated
uncertainties.

The predicted $M_{\ell b}$ distributions are calculated separately for
dilepton, single-tagged, and double-tagged data samples, by Monte Carlo
simulations of $t\bar{t}$ and background. The effects of predicted
kinematics, decay distributions, detector acceptance, and resolution are all
considered.  The HERWIG event generator~\cite{herwig} with the MRST h-g PDF
set~\cite{MRST} was used to model $t\bar{t}$ production.


%
For cases with two possible $b$ jets that can be matched to a
lepton (the dilepton and double-tagged samples), the fit is performed to
two-dimensional distributions of $M^2_{\ell b_{(1)}}$ and $M^2_{\ell
b_{(2)}}$, thus taking into account that only one can be correct.  Naively,
this ambiguity in assignments of leptons and $b$ quarks to one top quark
would appear to be problematic in this measurement.  However, while correct
pairings are limited kinematically by $M_t^2 - M_W^2$ for a massless $b$
quark, incorrect pairings often have significantly higher mass.  With our two
dimensional fit, mispairings only increase the statistical
uncertainty in the fit by only 15\%.

\begin{table}
  \begin{tabular}{|c|c|}
   \hline
   \multicolumn{2}{|c|}{Systematic Uncertainties}\\  
   \hline
   Top mass                        &    0.19  
                                              \\
   Jet energy scale                &    0.04  
                                              \\
   Background shape                &    0.05  \\
   Background normalization        &    0.05  \\
   ISR gluon radiation             &    0.04  \\
   FSR gluon radiation             &    0.03  \\
   B tagging efficiency            &    0.03  \\
   Parton distribution functions   &    0.02  \\
   Monte Carlo statistics          &    0.01  \\
   Relative acceptance             &    0.005  \\
   \hline
   Total  systematic  &     0.21     \\
   \hline
   \end{tabular}
\caption{Summary of systematic uncertainties in terms of the shift in
measurement of the V$+$A
  fraction.  The systematic uncertainties shown for the top mass and jet
energy scale are after considering the correlations between the two; without these corrections the systematic uncertainties are $0.21$ and $0.14$, respectively.}
\label{table:system}
\end{table}

Systematic uncertainties in the measurement enter the analysis primarily
through the prediction of the $M_{\ell b}$ distributions, and are evaluated
by changing assumptions in the Monte Carlo simulation.  Listed individually
in Table~\ref{table:system}, all systematic uncertainties added in quadrature
represent a $0.21$ uncertainty in $f_{V+A}$.  The largest systematic
uncertainties are from the top mass and the jet energy scale.  Increasing the
top mass will increase $M_{\ell b}$ in top decay.  The measured uncertainty
of the top quark mass is $5.1$~GeV~\cite{combination}, and an increase
in top mass by one standard deviation increases $f_{V+A}$ by $0.19$.  
Sources of systematic uncertainty in the jet energy scale include the
calibration of the calorimeter, the simulation of the calorimeter response
and the modeling of fragmentation~\cite{toplj}.  An increase in the overall
jet energy scale by one standard deviation would increase $f_{V+A}$ by
$0.14$.  However, the CDF jet energy scale has a large effect on the world
average top mass measurement.  Accounting for the correlation between
these two effects results in a reduction of the systematic from jet energy
scale to 0.04.

Smaller sources of systematic uncertainties were studied in
this measurement by observing the effect in simulated
pseudoexperiments.  
Hard gluon bremsstrahlung either in the
initial or final state can cause
significant mismeasurement of the $b$ quark jet or can produce a jet
which can be mistaken for the $b$ quark jet itself.  The size of the
effect was conservatively estimated by removing all such events from the
sample in a simulated measurement. 
For samples where SVX topological $b$ tagging was used, the effect of
uncertainties in $b$ tagging efficiency as a function of $b$ jet $E_T$
were evaluated.  
Estimated background rates and distributions in
$M^2_{\ell b}$ were varied as well.  The most important of these
effects is the uncertainty in the mean $Q^2$ used in the VECBOS simulation of
the $W$+jets background as discussed in Ref.~\cite{cdf_mass_prd}.
A set of CTEQ~\cite{cteq} and MRST~\cite{MRST} Parton Distribution Functions (PDFs)
were compared to the standard PDF set of MRST h-g and found to cause a
small spread in the measured $f_{V+A}$.
Systematic uncertainty due to the limited size of the Monte
Carlo simulation samples is also included.

%



\begin{figure*}
  \resizebox{0.3\textwidth}{!}{\includegraphics{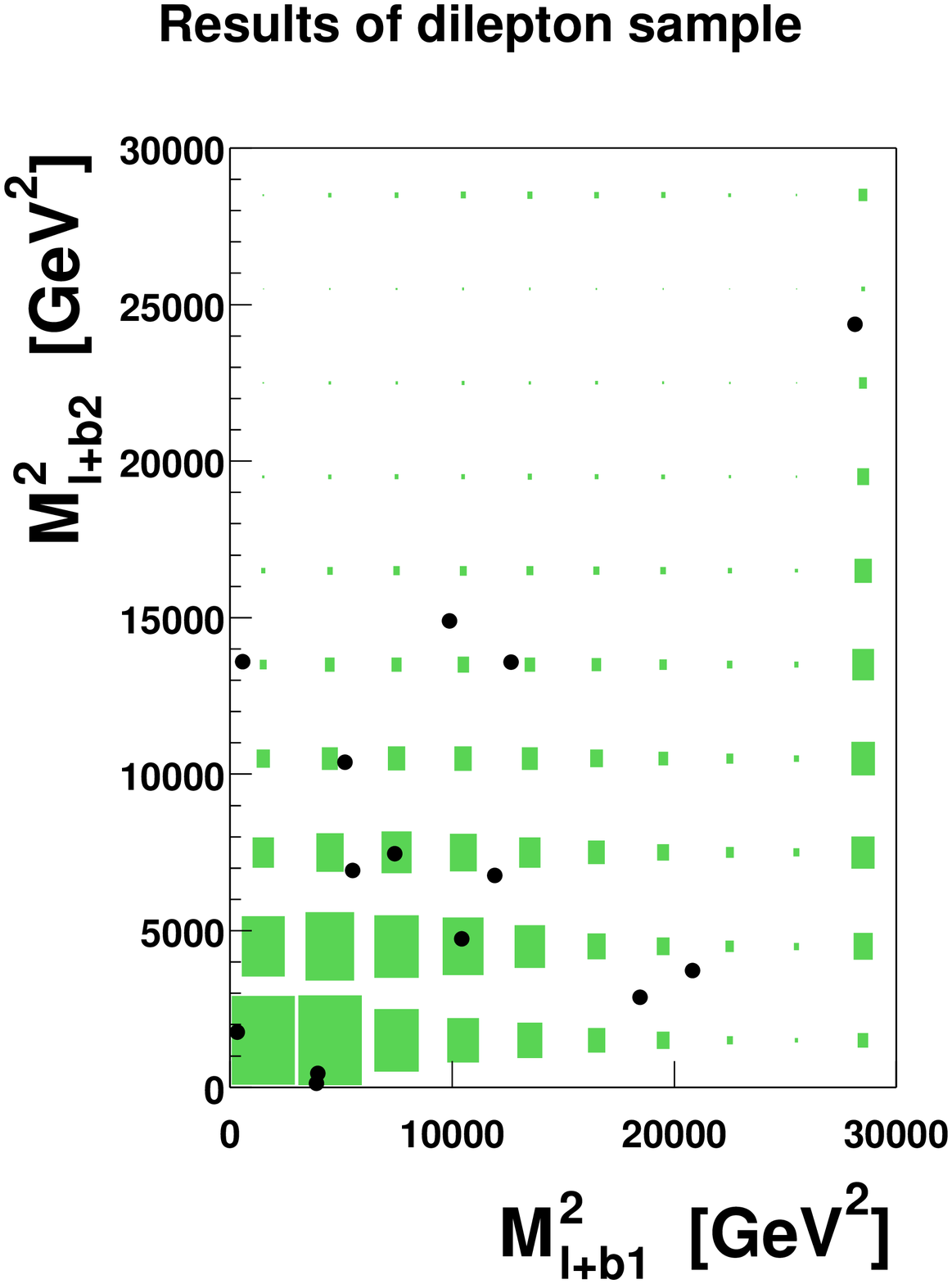}}
  \resizebox{0.3\textwidth}{!}{\includegraphics{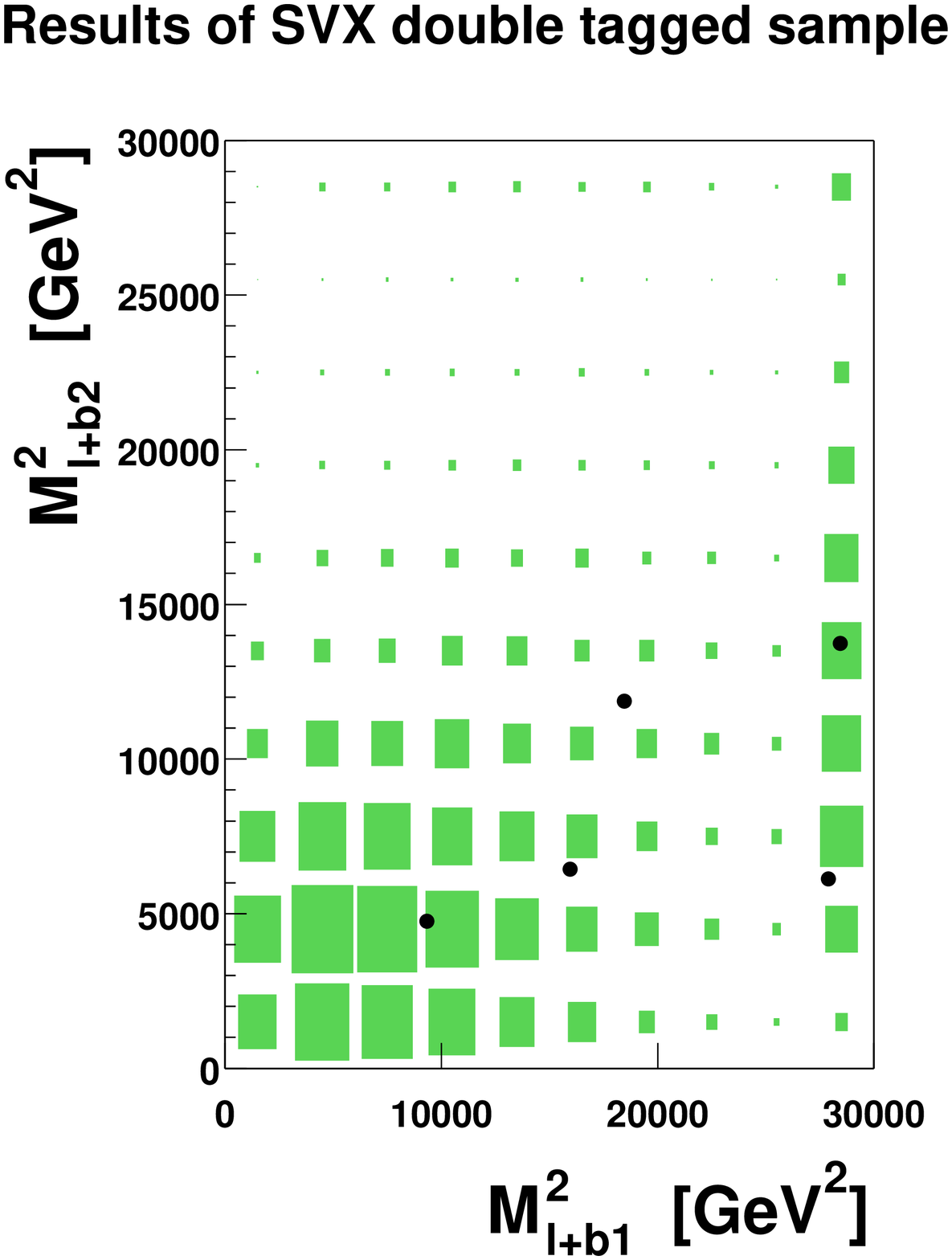}}
  \resizebox{0.3\textwidth}{!}{\includegraphics{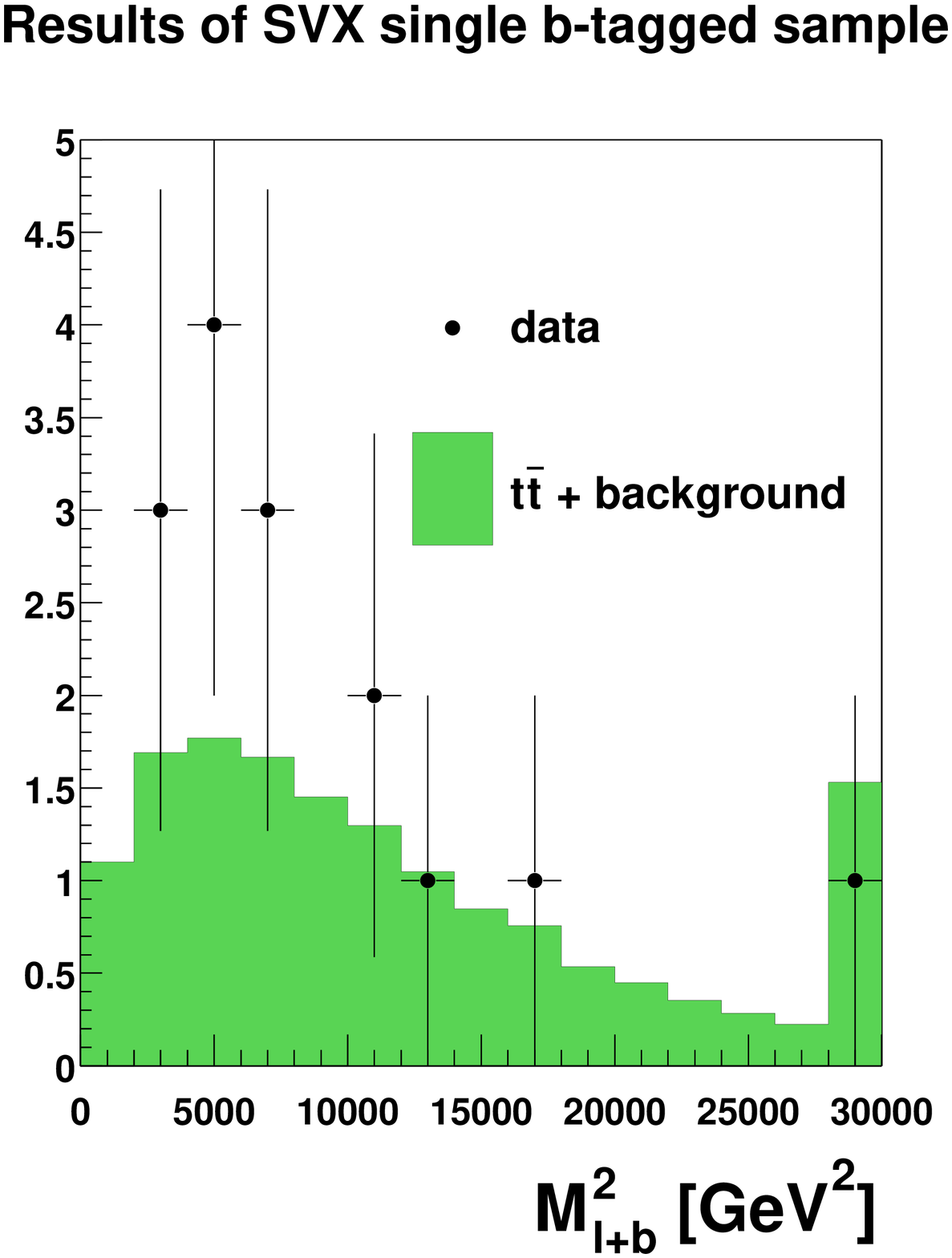}}
\caption{Data and Standard Model Monte Carlo distributions for each
sample. The last bin 
includes combinations greater than 30,000 $GeV^2$, which are predominantly
the result of incorrect pairings. Errors are statistical only.}
\label{fig:results}
\end{figure*}

\begin{figure}
\resizebox{0.9\columnwidth}{!}{\includegraphics{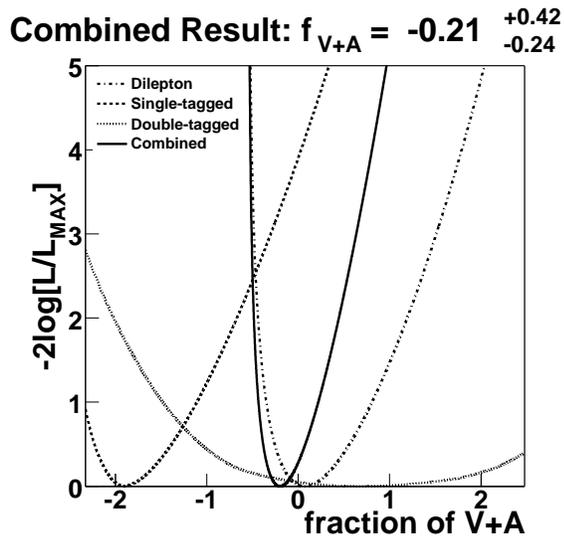}}
\caption{$-2\log{\cal L}$ as a function of $f_{V+A}$ for all samples
and for the combined likelihood fit.  The result for the dilepton sample 
is $f_{V+A}$ = 0.08$^{+0.74}_{-.0.42}$, for the single-tagged sample is
$f_{V+A}$ = -1.91$^{+0.69}_{-.0.48}$, and for the double-tagged sample is
$f_{V+A}$ = 0.63$^{+2.62}_{-2.11}$.  Errors are statistical only.}
\label{fig:combined_results}
\end{figure}

The data and expected Standard Model distributions are shown for each of the
three samples in Fig.~\ref{fig:results}.  We can combine
the statistical likelihood as a function of $f_{V+A}$ for each sample into
the joint likelihood shown in Fig.~\ref{fig:combined_results}.  The combined
result for $f_{V+A}$ and its $1\sigma$ uncertainties are 
\begin{equation}
f_{V+A} =
-0.21_{-0.24}^{+0.42}({\rmt stat.\,})\pm0.21({\rmt syst.\,})
\end{equation}

The central value depends on the true top mass, $f_{V+A} (M_t) = -0.21 +
0.037 (M_t - 174.3$ GeV), and the top mass uncertainty is reflected in the
systematic error.  This central value lies in an unphysical region, but is
more consistent with a Standard Model V$-$A interaction for the $tbW$ vertex
than a V$+$A interaction.  We can place a one-sided upper limit on the
fraction of rate due to a V$+$A component by construction of a Neyman
confidence band in the variable $f_{V+A}$ ~\cite{neyman}. This procedure
results in an upper limit on $f_{V+A}$ of $0.80$ at 95\% confidence level.
With the assumption of a standard model longitudinal helicity fraction, this
corresponds to $f_{+} < 0.24$ at 95\% confidence level.

$W$ polarization in top decays has also been studied at CDF in the same data
sample using the lepton $P_T$ \cite{lpt} as the observable to discriminate
between left-handed and right-handed $W$ bosons, under the assumption of a
fixed longitudinal helicity.  These two results have different selection
criteria, but share largely overlapping data sets.  In addition, the
observables themselves are weakly correlated, and a large fraction of the
systematic uncertainties are common.  Nevertheless, the overall statistical
correlation of the two results is only about $0.4$.  Under the simplifying
assumption of Gaussian uncertainties, the combined measurement using both the
$M_{\ell b}$ and lepton $P_T$ approaches is that the fraction of $W$
bosons produced in a V$+$A interaction is
\begin{equation}
f_{V+A}=-0.07\pm0.37({\rmt stat.\, \oplus syst.\,}).
\end{equation}
The combined upper limit is $f_{V+A} < 0.61$ at 95\% confidence level.  In
terms of the right-handed helicity fraction, this corresponds to $f_{+} <
0.18$ at 95\% confidence level.  The combined result is inconsistent with a
pure V$+$A theory at a confidence level equivalent to the probability of a
$2.7\sigma$ Gaussian statistical fluctation.

In conclusion, we have used the measurement of $M_{\ell b}$ in $t\bar{t}$
events to measure the polarization of $W$ bosons in top decay.  The results
are consistent with the V$-$A theory of the weak interaction.  The data are
used to set a limit on the fraction of top 
decays mediated by a V$+$A interaction.
This is the first result providing significant direct evidence against a pure
V+A theory of weak interactions in top decay;
it also provides the first significant limits on partial admixtures of a
V$+$A interaction with the expected V$-$A reaction.  

We thank the Fermilab staff and the technical staffs of the participating
institutions for their vital contributions.  This work was supported by the
U.S. Department of Energy and National Science Foundation; the Italian
Istituto Nazionale di Fisica Nucleare; the Ministry of Education, Culture,
Sports, Science and Technology of Japan; the Natural Sciences and Engineering
Research Council of Canada; the National Science Council of the Republic of
China; the Swiss National Science Foundation; the A.P. Sloan Foundation; the
Research Corporation; the Bundesministerium fuer Bildung und Forschung,
Germany; the Korean Science and Engineering Foundation and the Korean
Research Foundation; the Particle Physics and Astronomy Research Council and
the Royal Society, UK; the Russian Foundation for Basic Research; the
Comision Interministerial de Ciencia y Tecnologia, Spain; work supported in
part by the European Community's Human Potential Programme under contract
HPRN-CT-20002, Probe for New Physics; and this work was supported by Research
Fund of Istanbul University Project No. 1755/21122001.


\begin{thebibliography}{9}
\bibitem{eta}
In the CDF coordinate system, $\theta$ and $\phi$ are the polar and azimuthal
angles, respectively, with respect to the proton beam direction which defines
the z axis.  The pseudorapidity $\eta$ is defined as $- \ln{\tan \frac{\theta}{2}}$.
\end{thebibliography}
\end{document}